\pdfoutput=1

\documentclass[12pt,prd,aps,amssymb,amsmath,tightenlines,showpacs]{revtex4}
\usepackage{graphicx}

\newcommand{\cP}{\ensuremath{\mathcal{P}}}
\newcommand{\cT}{\ensuremath{\mathcal{T}}}
\newcommand{\cPT}{\ensuremath{\mathcal{PT}}}
\newcommand{\half}{\mbox{$\textstyle{\frac{1}{2}}$}}

\begin{document}

\title{Systems of coupled $\cPT$-symmetric oscillators}

\author{Carl M. Bender$^{a,b}$}\email{cmb@wustl.edu}
\author{Mariagiovanna Gianfreda$^{a,c}$}\email{Maria.Gianfreda@le.infn.it}
\author{S. P. Klevansky$^d$}\email{spk@physik.uni-heidelberg.de}

\affiliation{$^a$Department of Physics, Washington University, St. Louis, MO
63130, USA\\
$^b$Department of Mathematical Science, City University London,
Northampton Square, London EC1V 0HB, UK\\
$^c$Institute of Industrial Science, University of Tokyo, Komaba, Meguro, Tokyo
153-8505, Japan\\
$^d$Institut f\"ur Theoretische Physik, Universit\"at Heidelberg, Philosophenweg
19, 69120 Heidelberg, Germany\footnote{Permanent address.} and Department of
Physics, University of the Witwatersrand, Johannesburg, South Africa}

\date{\today}

\begin{abstract}
The Hamiltonian for a $\cPT$-symmetric chain of coupled oscillators is
constructed. It is shown that if the loss-gain parameter $\gamma$ is uniform for
all oscillators, then as the number of oscillators increases, the region of
unbroken $\cPT$-symmetry disappears entirely. However, if $\gamma$ is localized
in the sense that it decreases for more distant oscillators, then the
unbroken-$\cPT$-symmetric region persists even as the number of oscillators
approaches infinity. In the continuum limit the oscillator system is described
by a $\cPT$-symmetric pair of wave equations, and a localized loss-gain impurity
leads to a pseudo-bound state. It is also shown that a planar configuration of
coupled oscillators can have multiple disconnected regions of unbroken $\cPT$
symmetry.
\end{abstract}


\maketitle

\section{Introduction}
\label{s1}
A previous paper \cite{R1} considered a system consisting of a pair of coupled
oscillators, one with loss and the other with gain. Such a system is $\cPT$
symmetric if the loss and gain parameters are equal. The energy of this
$\cPT$-symmetric system is exactly conserved because this system is described by
a Hamiltonian. In the current paper we examine the systems that arise when the
number of pairs of coupled oscillators is extended from $1$ to $N$, where $N$
can be arbitrarily large.

Let us review the case $N=1$. A single pair of coupled oscillators, the first
with loss and the second with gain, is described by the equations of motion
\begin{equation}
\ddot{x}+\omega^2x+\mu\dot{x}=-\epsilon y,\quad
\ddot{y}+\omega^2y-\nu\dot{y}=-\epsilon x.
\label{E1}
\end{equation}
To treat this system at a classical level, we seek solutions to (\ref{E1}) of
the form $e^{i\lambda t}$. The classical frequency $\lambda$ then satisfies the
quartic polynomial equation 
\begin{eqnarray}
\lambda^4-i(\mu-\nu)\lambda^3-(2\omega^2-\mu\nu)\lambda^2+i\omega^2(\mu-\nu)
\lambda-\epsilon^2+\omega^4=0.
\label{E2}
\end{eqnarray}

This classical system becomes $\cPT$ symmetric if the loss and gain are
balanced; that is, if we set $\mu=\nu=2\gamma$. In this case the frequencies
$\lambda$ are given by
\begin{equation}
\lambda^2=\omega^2-2\gamma^2\pm\sqrt{\epsilon^2-4\gamma^2\omega^2+4\gamma^4}.
\label{E3}
\end{equation}
Note that there are four real frequencies when $\epsilon$ is in the range
\begin{equation}
\epsilon_1=2\gamma\sqrt{\omega^2-\gamma^2}<\epsilon<\epsilon_2=\omega^2.
\label{E4}
\end{equation}
This defines the {\it unbroken} $\cPT$-symmetric region. In the
broken-$\cPT$-symmetric region $\epsilon<\epsilon_1$ there are two pairs of
complex-conjugate frequencies and in the broken-$\cPT$-symmetric region
$\epsilon>\epsilon_2$ there are two real frequencies and one complex-conjugate
pair of frequencies.

When $\mu\neq\nu$, the system (\ref{E1}) is not Hamiltonian. However, when the
system is $\cPT$ symmetric ($\mu=\nu=2\gamma$), (\ref{E1}) can be derived from
the two-coupled-oscillator Hamiltonian
\begin{equation}
H_2=pq+\gamma(yq-xp)+\left(\omega^2-\gamma^2\right)xy+\half\epsilon\left(x^2+
y^2\right).
\label{E5}
\end{equation}
This Hamiltonian is $\cPT$ symmetric because under parity reflection $\cP$
the loss and gain oscillators are interchanged \cite{R2},
\begin{equation}
\cP:~x\to-y,\quad y\to-x,\quad p\to-q,\quad q\to-p,
\label{E6}
\end{equation}
and under time reversal $\cT$ the signs of the momenta are reversed,
\begin{equation}
\cT:~x\to x,\quad y\to y,\quad p\to-p,\quad q\to-q.
\label{E7}
\end{equation}
The Hamiltonian $H_2$ is $\cPT$ symmetric but it is not invariant under $\cP$ or
$\cT$ separately \cite{R3}. Because the balanced-loss-gain system is
Hamiltonian, the energy (that is, the value of $H_2$) is conserved. However, the
total energy (\ref{E5}) is not the usual sum of kinetic and potential energies
(such as $p^2+q^2+x^2+y^2$).

If we set the coupling parameter $\epsilon$ to zero, $H_2$ describes the system
studied by Bateman \cite{R4}. Bateman showed that an equation of motion having a
friction term linear in velocity could be derived from a variational principle.
To do this he introduced a time-reversed companion of the original damped
harmonic oscillator. This auxiliary oscillator acts as an energy reservoir and
can be viewed as a thermal bath. The classical Hamiltonian for the Bateman
system was constructed by Morse and Feschbach \cite{R5} and the corresponding
quantum theory was analyzed by many authors, including Bopp \cite{R6}, Feshbach
and Tikochinsky \cite{R7}, Tikochinsky \cite{R8}, Dekker \cite{R9}, Celeghini,
Rasetti, and Vitiello \cite{R10}, Banerjee and Mukherjee \cite{R11}, and
Chru\'sci\'nski and Jurkowski \cite{R12}. Only the {\it noninteracting}
($\epsilon=0$) case was considered in these references.

The noteworthy feature of $\cPT$-symmetric systems is that they exhibit
transitions; the classical system described by $H_2$ exhibits two transitions.
The first occurs at $\epsilon=\epsilon_1$. If $\epsilon<\epsilon_1$, the energy
flowing into the $y$ resonator cannot transfer fast enough to the $x$ resonator,
where energy is flowing out, so the system cannot be in equilibrium. However,
when $\epsilon>\epsilon_1$, the energy flowing into the $y$ resonator transfers
to the $x$ resonator and the entire system is in equilibrium. The frequencies of
a classical system in equilibrium are real and the system exhibits {\it Rabi
oscillations} (power oscillations between the two resonators) in which the two
oscillators are $90^\circ$ out of phase. Complex frequencies indicate
exponential growth and decay and are a signal that the system is not in
equilibrium. A second transition occurs at $\epsilon=\epsilon_2$; when
$\epsilon>\epsilon_2$, the classical system is no longer in equilibrium. This
transition is difficult to see in classical experiments because in the
strong-coupling regime the loss and gain components would have to be so close
that they would interfere with one another. For example, in the pendulum
experiment in Ref.~\cite{R13} the pendula would be so close that they could no
longer swing freely, and in the optical-resonator experiment in Ref.~\cite{R14}
the solid-state resonators would be damaged. This strong-coupling region is
discussed for the case of coupled systems {\it without} loss and gain in
Ref.~\cite{R15}, where it is called the {\it ultrastrong-coupling regime}. 

In Ref.~\cite{R1} it is shown that the classical and the quantum systems
described by $H_2$ exhibit transitions at the same two values of the coupling
parameter $\epsilon$. When $\epsilon<\epsilon_1$ and when $\epsilon>\epsilon_2$
the quantum energies are complex, but in the unbroken-$\cPT$-symmetric region
$\epsilon_1<\epsilon<\epsilon_2$ the quantum energies are real.

This paper is organized as follows. In Sec.~\ref{s2} we formulate the equations
of motion for a linear chain of $N$ identical pairs of $\cPT$-symmetric
loss-gain oscillators and we construct the Hamiltonians $H_{2N}$ for such
systems. We show that there are two ways to represent such Hamiltonians, one
that we call a {\it sum} representation and another that we call a {\it product}
representation. In the product representation it is easy to see that the
Hamiltonian is not unique and that this nonuniqueness takes the form of a gauge
invariance. Next, in Sec.~\ref{s3} we construct the Hamiltonians for a general
$\cPT$-symmetric system of $2N$ coupled oscillators in which the coupling
parameter $\epsilon$ and the loss-gain parameter $\gamma$ are allowed to vary
from oscillator to oscillator. In addition, we consider a system of $2N+1$
coupled $\cPT$-symmetric oscillators, where $\cPT$ symmetry requires that the
central oscillator have neither loss nor gain. We also perform the $N\to\infty$
limit of $H_{2N}$. In this limit the equations of motion of the oscillators
become coupled linear wave equations with balanced loss and gain.

In Sec.~\ref{s4} we ask whether a $\cPT$-symmetric chain of $2N$ coupled
oscillators can have an unbroken-$\cPT$-symmetric region. We show that as $N$
increases, if $\gamma$ and $\epsilon$ are the same for all oscillators, the
region of unbroken $\cPT$ symmetry shrinks and disappears entirely as $N\to
\infty$. However, if the loss-gain parameter $\gamma$ decreases to 0 for distant
oscillators, then such systems always have an unbroken-$\cPT$-symmetric region
for intermediate values of the coupling parameter $\epsilon$ surrounded by
broken-$\cPT$-symmetric regions for small and large values of $\epsilon$.
Specifically, for the cases in which $\gamma_n$ decreases like $1/n$ or $1/n^2$,
where $1\leq n\leq N$ is the number of the oscillator measured from the center
of the system, we show that an unbroken-$\cPT$-symmetric region persists in the
limit as $N\to\infty$. If one views loss-gain as the consequence of an impurity,
then a configuration of oscillators for which $\gamma$ decreases with increasing
distance from the center can be seen as having a localized impurity. Thus, in
Sec.~\ref{s5} we investigate a special case for the continuum model in which
there is a point-like $\cPT$-symmetric impurity localized at the origin. We find
that this impurity gives rise to a pseudobound-state solution. In Sec.~\ref{s6}
we consider the simplest case of a two-dimensional array of coupled oscillators,
namely three oscillators, one with loss, one with gain, and the third with
neither loss nor gain. This system is interesting because it can exhibit {\it
five} distinct regions as a function of the coupling constant, two having
unbroken $\cPT$ symmetry and three having broken $\cPT$ symmetry. Finally, in
Sec.~\ref{s7} we make some brief concluding remarks.

\section{$\cPT$-symmetric system of coupled classical oscillators}
\label{s2}
In this section we describe the properties of a $\cPT$-symmetric one-dimensional
chain of $2N$ coupled oscillators with alternating loss and gain. We begin by
making the simplifying assumptions that the natural frequency $\omega$, the
coupling to adjacent oscillators $\epsilon$, and the loss-gain parameter
$\gamma$ are the same for all oscillators. The classical coordinates are
$x_k(t)$ ($1\leq k\leq 2N$) and the equations of motion are
\begin{eqnarray}
\ddot{x}_1+\omega^2x_1+2\gamma\dot{x}_1&=&-\epsilon x_2,\nonumber\\
\ddot{x}_2+\omega^2x_2-2\gamma\dot{x}_2&=&-\epsilon x_1-\epsilon x_3,\nonumber\\
\ddot{x}_3+\omega^2x_3+2\gamma\dot{x}_3&=&-\epsilon x_2-\epsilon x_4,\nonumber\\
\ddot{x}_4+\omega^2x_4-2\gamma\dot{x}_4&=&-\epsilon x_3-\epsilon x_5,\nonumber\\
\ldots&=&\ldots,\nonumber\\
\ddot{x}_{2N}+\omega^2x_{2N}-2\gamma\dot{x}_{2N}&=&-\epsilon x_{2N-1}.
\label{E8}
\end{eqnarray}
These equations of motion are $\cPT$ symmetric, where the definitions of $\cP$
and $\cT$ are generalized from (\ref{E6}) and (\ref{E7}) to
\begin{eqnarray}
\cP:&~&x_k\to-x_{2N-k+1},\quad p_k\to-p_{2N-k+1}\quad(1\leq k\leq2N),\nonumber\\
\cT:&~&x_k\to x_k,\quad p_k\to-p_k\quad(1\leq k\leq2N).
\label{E9}
\end{eqnarray}

The equations of motion (\ref{E8}) imply that there is a conserved quantity. To
construct this constant of the motion we multiply the first equation by
$\dot{x}_2$, the second equation by $\dot{x}_1+\dot{x}_3$, the third equation by
$\dot{x}_2+\dot{x}_4$, the fourth equation by $\dot{x}_3+\dot{x}_5$, and so on.
If we add the resulting equations, $\gamma$ drops out entirely and we obtain a
time-independent quantity, which we can identify as the energy $E_{2N}$ of the
system:
\begin{eqnarray}
E_{2N}=\sum_{j=1}^{2N-1}\left(\dot{x}_j\dot{x}_{j+1}+\omega^2 x_jx_{j+1}\right)
+\frac{\epsilon}{2}\left(x_1^2+x_{2N}^2\right)+\epsilon\sum_{j=2}^{2N-1}x_j^2
+\epsilon\sum_{j=1}^{2N-2}x_jx_{j+2}.
\label{E10}
\end{eqnarray}

The existence of a conserved quantity suggests that (\ref{E8}) is a Hamiltonian
system, and indeed one can find a Hamiltonian from which these equations of
motion can be derived. There are two ways to express the (nonunique) Hamiltonian
that gives rise to (\ref{E8}); we can use what we call a {\it sum} or a {\it
product} representation. We describe these two structures below.

\subsection{Sum representation of the Hamiltonian}
\label{ss2A}
In the sum representation $H_{2N}$ consists of four terms. First, there is a
pure momentum term of the form $p_1p_2+p_2p_3+p_3p_4+\ldots+p_{2N-1}p_{2N}$.
Second, there is a momentum times a coordinate term proportional to $\gamma$:
$\gamma\left(-p_1x_1+p_2x_2-p_3x_3+\ldots+p_{2N}x_{2N} \right)$. Third, there is
a potential-energy-like term proportional to $\epsilon$: $\half\epsilon\left(
x_1^2+x_2^2+x_3^2+\ldots+x_{2N}^2\right)$. (It is surprising that this term is
proportional to $\epsilon$ because in the equations of motion $\epsilon$ appears
to play the role of a coupling constant; $\epsilon$ does not appear to be a
measure of the potential energy, which one associates with a frequency of
oscillation.) Fourth, there is an oscillator coupling term proportional to
$\omega^2-\gamma^2$:
\begin{eqnarray}
\big[x_1x_2+x_3x_4+x_5x_6+x_7x_8+\hspace{-1.0cm}&\ldots&\hspace{-1.0cm}
+x_{2N-7}x_{2N-6}+x_{2N-5}x_{2N-4}+x_{2N-3}x_{2N-2}+x_{2N-1}x_{2N}\nonumber\\
-x_1x_4-x_3x_6-x_5x_8-\hspace{-1.0cm}&\ldots&\hspace{-1.0cm}-x_{2N-7}x_{2N-4}
-x_{2N-5}x_{2N-2}-x_{2N-3}x_{2N}\nonumber\\
+x_1x_6+x_3x_8+\hspace{-1.0cm}&\ldots&\hspace{-1.0cm}+x_{2N-7}x_{2N-2}
+x_{2N-5}x_{2N}\nonumber\\
-x_1x_8-\hspace{-1.0cm}&\ldots&\hspace{-1.0cm}-x_{2N-7}x_{2N}\nonumber\\
&\ldots&\nonumber\\
&(-1)^{N+1}x_1x_{2N}&\big]\big(\omega^2-\gamma^2\big).
\label{E11}
\end{eqnarray}
Note the interesting structure of this term: The jumps in the products in
(\ref{E11}) skip 0, 2, 4, 6, ... and change sign. A compact expression for
$H_{2N}$ is
\begin{eqnarray}
H_{2N}=\sum_{j=1}^{2N-1}p_jp_{j+1}+\frac{\epsilon}{2}\sum_{j=1}^{2N}x_j^2+
\gamma\sum_{j=1}^{2N}(-1)^jx_jp_j+\left(\omega^2-\gamma^2\right)\sum_{j=0}^{N-1}
(-1)^j\sum_{k=1}^{N-j}x_{2k-1}x_{2j+2k}.
\label{E12}
\end{eqnarray}

To obtain the equations of motion (\ref{E8}) for this Hamiltonian from
Hamilton's equations, we take one derivative of $H_{2N}$ with respect to $p_k$
to find $\dot{x}_k$:
\begin{equation}
\dot{x}_k=p_{k+1}+p_{k-1}+(-1)^k \gamma x_k.
\label{E12.5}
\end{equation}
We then take a time derivative,
\begin{eqnarray}
\ddot{x}_k-(-1)^k\gamma\dot{x}_k&=& -\frac{\partial H_{2N}}{\partial x_{k+1}}
-\frac{\partial H_{2N}}{\partial x_{k-1}}\nonumber\\
&=&-\epsilon x_{k+1}-\epsilon x_{k-1}+(-1)^k\gamma\left(p_{k+1}+p_{k-1}\right)
+\left(\omega^2-\gamma^2\right)(\ldots),
\label{E13}
\end{eqnarray}
and use the one-derivative equation (\ref{E12.5}) to recover the equations of
motion (\ref{E8}).

\subsection{Product representation of the Hamiltonian}
\label{ss2B}
In this representation it is easy to understand the nonuniqueness of the
Hamiltonian that gives rise to the equations of motion (\ref{E8}). This
nonuniqueness is a gauge invariance, where $\gamma$ plays the role of an
electric charge. Without changing the equations of motion we rewrite the sum
representation $H_2$ in (\ref{E5}) so that the momentum terms appear in factored
form:
\begin{equation}
H_2=(p+\gamma y)(q-\gamma x)+\omega^2 xy+\epsilon(x^2+y^2)/2.
\label{E14}
\end{equation}
Similarly, the sum representation for $H_4$,
\begin{eqnarray}
H_4&=& p_1p_2+p_2p_3+p_3p_4+\epsilon(x_1^2+x_2^2+x_3^2+x_4^2)/2\nonumber\\
&&+\gamma(-x_1p_1+x_2p_2-x_3p_3+x_4p_4)+(\omega^2-\gamma^2)(x_1x_2+x_3x_4-x_1
x_4)
\label{E15}
\end{eqnarray}
can be reconfigured in product form as
\begin{eqnarray}
H_4&=&[p_1+\gamma(x_2-x_4)](p_2-\gamma x_1)+(p_2-\gamma x_1)(p_3+\gamma x_4)
+(p_3+\gamma x_4)[p_4-\gamma(x_3-x_1)]\nonumber\\
&&+\omega^2(x_1x_2+x_3x_4-x_1x_4)+\epsilon(x_1^2+x_2^2+x_3^2+x_4^2)/2
\label{E16}
\end{eqnarray}
without changing the equations of motion. The product representation of $H_6$
has the form
\begin{eqnarray}
H_6&=&[p_1+\gamma(x_2-x_4+x_6)](p_2-\gamma x_1)+(p_2-\gamma x_1)[p_3+\gamma(x_4
-x_6)]\nonumber\\
&&+[p_3+\gamma(x_4-x_6)][p_4-\gamma(x_3-x_1)]+[p_4-\gamma(x3-x_1)](p_5+\gamma
x_6)\nonumber\\
&&+(p_5+\gamma x_6)[p_6-\gamma(x_5-x_3+x_1]+\omega^2(x_1y_1+x_3x_4+x_5x_6-x_1x_4
-x_3x_6+x_1x_6)\nonumber\\
&&+\epsilon(x_1^2+x_2^2 + x_3^2 +x_4^2 + x_5^2 +x_6^2)/2.
\label{E17}
\end{eqnarray}
The general structure for the product representation of $H_{2N}$ is now clear.

The advantage of the product representation is that if we consider the
Hamiltonian to be quantum mechanical, we can identify a gauge invariance. Each
momentum factor in the product representation has the form $[p+\gamma({\rm sum~
of~spatial~coordinates})]$. This term resembles the structure $p-eA$ in
electrodynamics, which suggests that we can make a unitary (canonical)
transformation analogous to a gauge transformation in electrodynamics. By virtue
of the Heisenberg algebra $[x,p]=i$, it follows that $e^{-iax}pe^{iax}=p+a$,
where $a$ is a constant. Therefore, if we perform the unitary transformation 
\begin{equation}
e^{-ia_{mn}x_mx_n}H_{2N}e^{ia_{mn}x_mx_n}
\label{E18}
\end{equation}
on the Hamiltonian, the only terms that will be affected are the product terms
because they contain the momentum operators. The only changes that will occur
are that the momentum operators $p_m$ and $p_n$ will be shifted by terms that
are linear in the coordinates $x_n$ and $x_m$. There are $N(2N-1)$ independent
gauge transformations that can be performed on $H_{2N}$, and therefore we can
introduce $N(2N-1)$ arbitrary constants $a_{mn}$ into $H_{2N}$. Furthermore,
since the transformation is unitary, it leaves the equations of motion invariant
\cite{R16}.

\subsection{Lagrangian}
\label{ss2C}
Having found a Hamiltonian for the system (\ref{E8}), it is easy
to construct a Lagrangian:
\begin{eqnarray}
L_{2N}&=&\sum_{j=0}^{N-1}(-1)^j\sum_{k=1}^{N-j}\big[\gamma\left(\dot{x}_{2k-1}
x_{2j+2k}-x_{2k-1}\dot{x}_{2j+2k}\right)\nonumber\\
&&-\omega^2x_{2k-1}x_{2j+2k}+\dot{x}_{2k-1}\dot{x}_{2j+2k}\big]-\frac{\epsilon}
{2}\sum_{j=1}^{2N}x_j^2.
\label{E19}
\end{eqnarray}

\section{General case of nonconstant $\epsilon$, $\gamma$, $\omega$}
\label{s3}
We can construct a Hamiltonian (in the sum representation) for a
$\cPT$-symmetric system of $2N$ oscillators even if the parameters $\epsilon$,
$\gamma$, and $\omega$ vary from oscillator to oscillator
\begin{eqnarray}
H_{2N}&=&\sum_{k=1}^N(-1)^k\gamma_k\left(x_kp_k-x_{2N+1-k}p_{2N+1-k}\right)
+\sum_{k=1}^{N-1}\epsilon_k\left(x_kx_{2N-k}+x_{k+1}x_{2N+1-k}\right)
\nonumber\\
&&+\epsilon_N\left(x_N^2+x_{N+1}^2\right)/2+\sum_{k=1}^Np_kp_{2N+1-k}
+\sum_{k=1}^N\left(\omega_k^2-\gamma_k^2\right)x_kx_{2N+1-k}.
\label{E20}
\end{eqnarray}
We can also construct a Hamiltonian for $2N+1$ oscillators:
\begin{eqnarray}
H_{2N+1}&=&\sum_{k=1}^N(-1)^k\gamma_k\left(x_kp_k-x_{2N+2-k}p_{2N+2-k}\right)
+\sum_{k=1}^N\epsilon_k\left(x_kx_{2N+1-k}+x_{k+1}x_{2N+2-k}\right)\nonumber\\
&&+\left(x_{N+1}^2+p_{N+1}^2\right)/2+\sum_{k=1}^Np_kp_{2N+2-k}
+\sum_{k=1}^N\left(\omega_k^2-\gamma_k^2\right)x_kx_{2N+2-k}.
\label{E21}
\end{eqnarray}
The even Hamiltonian $H_{2N}$ leads to the equations of motion
\begin{eqnarray}
\ddot{x}_1+\omega_1^2x_1+2\gamma_1\dot{x}_1&=&-\epsilon_1x_2,\nonumber\\
\ddot{x}_2+\omega_2^2x_2-2\gamma_2\dot{x}_2&=&-\epsilon_1x_1-\epsilon_2x_3,
\nonumber\\
&\ldots&\nonumber\\
\ddot{x}_N+\omega_N^2x_N-(-1)^N2\gamma_N\dot{x}_N&=&
-\epsilon_{N-1}x_{N-1}-\epsilon_Nx_{N+1},\nonumber\\
\ddot{x}_{N+1}+\omega_N^2x_{N+1}+(-1)^N2\gamma_N\dot{x}_{N+1}&=&
-\epsilon_Nx_N-\epsilon_{N-2}x_{N+2},\nonumber\\
&\ldots&\nonumber\\
\ddot{x}_{2N-1}+\omega_2^2x_{2N-1}+2\gamma_2\dot{x}_{2N-1}&=&-\epsilon_1x_{2N}
-\epsilon_2x_{2N-2},\nonumber\\
\ddot{x}_{2N}+\omega_1^2x_{2N}-2\gamma_1\dot{x}_{2N}&=& -\epsilon_1x_{2N-1},
\label{E22}
\end{eqnarray}
and the odd Hamiltonian $H_{2N+1}$ gives the equations of motion
\begin{eqnarray}
\ddot{x}_1+\omega_1^2x_1+2\gamma_1\dot{x}_1&=&-\epsilon_1x_2,\nonumber\\
\ddot{x}_2+\omega_2^2x_2-2\gamma_2\dot{x}_2&=&-\epsilon_1x_1-\epsilon_2x_3,
\nonumber\\
&\ldots&\nonumber\\
\ddot{x}_{N+1}+\omega_{N+1}^2x_{N+1}&=&-\epsilon_N\left(x_N+x_{N+2}\right),
\nonumber\\
&\ldots&\nonumber\\
\ddot{x}_{2N}+\omega_2^2x_{2N}+2\gamma_2\dot{x}_{2N}&=&-\epsilon_1x_{2N+1}-
\epsilon_2x_{2N-1},\nonumber\\
\ddot{x}_{2N+1}+\omega_1^2x_{2N+1}-2\gamma_1\dot{x}_{2N+1}&=&-\epsilon_1x_{2N}.
\label{E23}
\end{eqnarray}

\subsection{Continuum limit $N\to\infty$}
\label{ss3A}
\begin{figure}[h!]
\begin{center}
\includegraphics[scale=0.60]{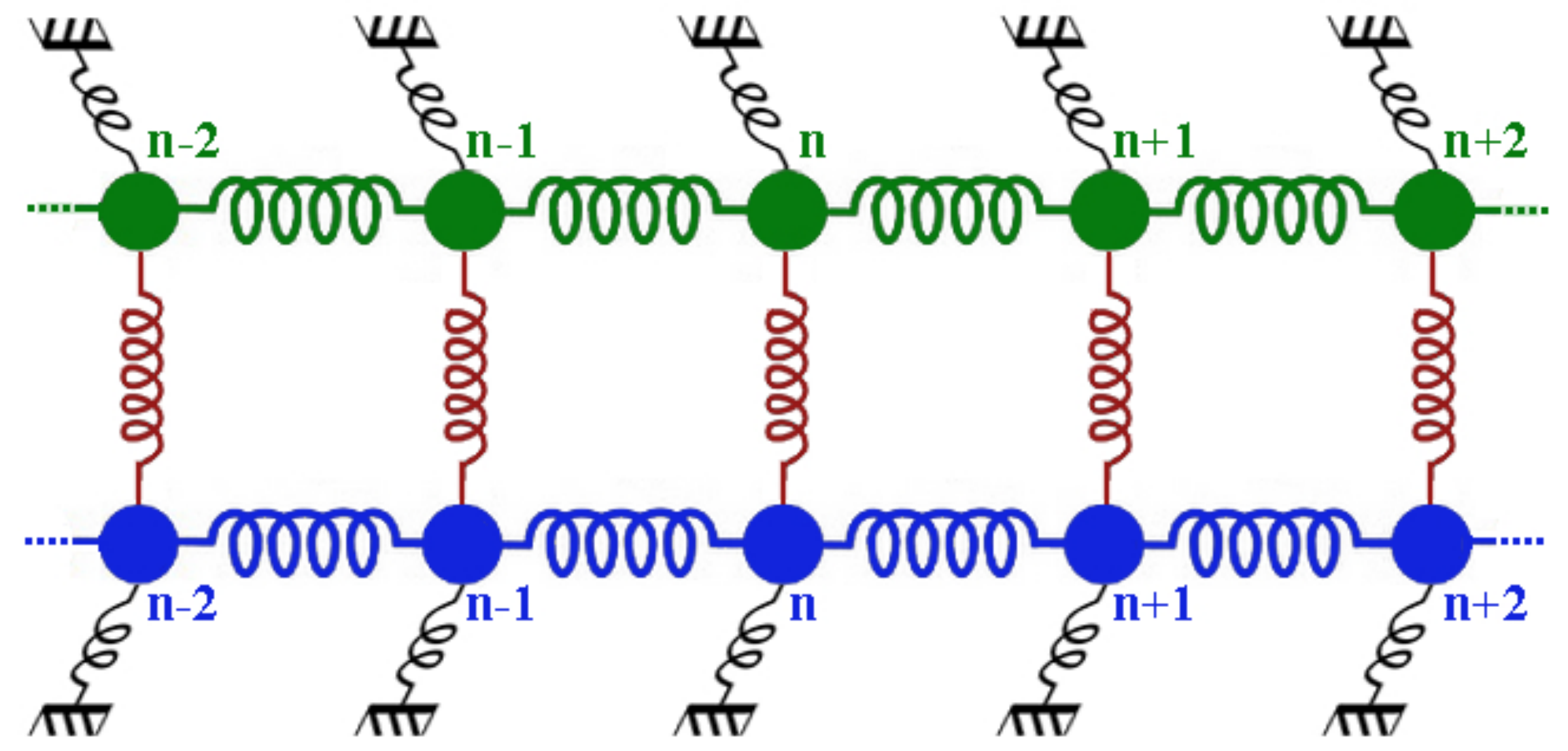}
\end{center}
\caption{Infinite $\cPT$-symmetric array of identical particles coupled by
springs. The masses in the top row, whose position coordinates are $x_n(t)$,
experience loss and the masses in the bottom row, which are located at $y_n(t)$,
experience gain.}
\label{F1}
\end{figure}

In this subsection we show how to take the limit as the number of oscillators
approaches infinity. For simplicity, let us consider two rows of identical
particles of mass $m$. These masses are coupled by springs, as illustrated in
Fig.~\ref{F1}.

The top row of particles is subject to damping (friction) forces and the bottom
row is subject to undamping forces. Each particle in the top row is coupled by
horizontal springs (of force constant per unit length $k/\Delta$) to the
adjacent particles to the left and right. Thus, the particle at $x_n$ is coupled
to its neighbors at $x_{n-1}$ and at $x_{n+1}$. The neighboring particles exert
a net force on the $n$th mass of strength $\frac{k}{\Delta}\big(x_{n+1}-2x_n+
x_{n-1}\big)$, where $\Delta$ is the equilibrium spacing. The constant $k$ is
the {\it tension} in the horizontal chain of masses. Also, there are fixed
springs above the top row of masses that exert a restoring force per unit length
of $-\mu_1\nu_1^2\Delta$ on each of the $x$ masses. This force tends to pull the
$x$ masses back to their equilibrium positions. The parameter $\mu_1$ has
dimensions of mass density (mass per unit length) and the parameter $\nu_1$ is a
frequency having dimensions of $1/{\rm time}$. The force on the $n$th mass due
to these vertical springs is $-\mu_1\nu_1^2\Delta x_n$. Finally, the particle at
$x_n$ in the top row is coupled to the particle at the position $y_n$ in the
bottom row by a vertical spring of force per unit length $\mu_2 \nu_2^2\Delta$.
(Here, $\mu_2$ is a mass density and $\nu_2$ is a frequency.) The force exerted
on the mass at $x_n$ due to the particle at $y_n$ is $\mu_2\nu_2^2\Delta\left(
y_n-x_n\right)$. The particles in the top row lose energy due to friction
(drag), where the dissipation per unit length is given by $\Gamma$. Thus, the
equation of motion of the $n$th particle is
\begin{equation}
m{\ddot x}_n+\Gamma\Delta{\dot x}_n=\frac{k}{\Delta}\left(x_{n+1}-2x_n+x_{n-1}
\right)-\mu_1\nu_1^2\Delta x_n+\mu_2\nu_2^2\Delta\left(y_n-x_n\right).
\label{E24}
\end{equation}

Let $m=\rho\Delta$, where $\rho$ is the horizontal mass per unit length. We then
divide (\ref{E24}) by $\Delta$ and take the limit as $\Delta\to0$ to get the
continuum wave equation
\begin{equation}
\rho u_{tt}+\Gamma u_t=k u_{xx}-\mu_1\nu_1^2u+\mu_2\nu_2^2(v-u).
\label{E25}
\end{equation}
Finally, we divide by $\rho$ and define the quantities $c^2\equiv k/\rho$,
$\gamma\equiv\Gamma/\rho$, $\omega^2\equiv\big(\mu_1\nu_1^2+\mu_2\nu_2^2\big)/
\rho$, and $\quad\epsilon\equiv-\mu_2\nu_2^2/\rho$. This leads to the wave
equation
\begin{equation}
u_{tt}+2\gamma u_t+\omega^2 u-c^2u_{xx}=-\epsilon v.
\label{E27}
\end{equation}
Similarly, from the equation for the particle at $y_n$ we obtain the wave
equation
\begin{equation}
v_{tt}-2\gamma v_t+\omega^2 v -c^2v_{xx}=-\epsilon u.
\label{E28}
\end{equation}
These equations are the continuous analogs of (\ref{E8}).

In anticipation of the calculation in Sec.~\ref{s5}, we rewrite these equations
in a more convenient form by defining $S(x,t)\equiv u(x,t)+v(x,t)$ and $D(x,t)
\equiv u(x,t)-v(x,t)$. The coupled wave equations satisfied by $S$ and $D$ are
\begin{eqnarray}
S_{tt}+\omega^2S-c^2S_{xx}+\epsilon S&=&-2\gamma(x)D_t,\nonumber\\
D_{tt}+\omega^2D-c^2D_{xx}-\epsilon D&=&-2\gamma(x)S_t,
\label{E30}
\end{eqnarray}
where we have now taken the loss-gain parameter $\gamma$ to depend on $x$.

\section{Existence of an unbroken-$\cPT$-symmetric region}
\label{s4}
The question addressed in this section is whether a region of unbroken $\cPT$
symmetry persists as the number of oscillators $N$ increases. We consider first
the case in which the loss-gain parameter $\gamma$ is the same for all
oscillators and show that the unbroken region disappears as $N$ increases. Next,
we demonstrate numerically that if $\gamma$ decreases for the more distant
oscillators, a region of unbroken $\cPT$ symmetry persists as $N\to\infty$.

\subsection{Case of constant $\gamma$}
\label{ss4A}
To find the frequencies of the system (\ref{E8}), we seek solutions of the form
$x_k=A_ke^{i\lambda t}$. The frequencies $\lambda$ can then be found by imposing
the condition that $\det\left[M_{2N}\right]=0$ (Cramer's rule), where $M_{2N}$ is the
$2N\times2N$ tridiagonal matrix
\begin{equation}
M_{2N}=\left(\begin{array}{ccccccc}
a-i b & -\epsilon & 0 & 0 & 0&0&\dots \\
-\epsilon & a+i b & -\epsilon & 0 & 0 &0& \dots\\
0 & -\epsilon & a-i b & -\epsilon & 0 & 0&\dots \\
0 & 0 & -\epsilon & a+ib & -\epsilon & 0&\dots\\
0 & 0 & 0 & -\epsilon & a-ib & -\epsilon & \dots \\
0 & 0 & 0 & 0& -\epsilon & a+ib &\dots\\
\vdots & \vdots & \vdots & \vdots & \vdots & \vdots & \ddots
\end{array}\right)
\label{E31}
\end{equation}
and $a$ and $b$ are given by $a=\lambda^2-\omega^2$ and $b=2\lambda\gamma$.

Let $P_N=\det\left[M_{2N}\right]$ ($N=1,2,\ldots$) be the polynomial obtained
by computing the determinant of the matrix $M_{2N}$. The first five of these
polynomials are
\begin{eqnarray}
P_1&=&-\epsilon^2+x,\nonumber\\
P_2&=&\epsilon^4-3x\epsilon^2+x^2,\nonumber\\
P_3&=&-\epsilon^6+6x\epsilon^4-5x^2\epsilon^2+x^3,\nonumber\\
P_4&=&\epsilon^8-10x\epsilon^6+15x^2\epsilon^4-7x^3\epsilon^2+x^4,\nonumber\\
P_5&=&-\epsilon^{10}+15x\epsilon^8-35x^2\epsilon^6+28x^3\epsilon^4-9x^4
\epsilon^2+x^5,
\label{E32}
\end{eqnarray}
where $x=a^2+b^2=\lambda^4+\lambda^2(4\gamma^2-2\omega^2)+\omega^4$. These
polynomials satisfy the recursion relation
\begin{equation}
P_N=\big(x-2\epsilon^2\big)P_{N-1}-\epsilon^4 P_{N-2}\quad(N\geq2),
\label{E33}
\end{equation}
where we take $P_0=1$.

\begin{figure}[h!]
\begin{center}
\includegraphics[scale=1.20]{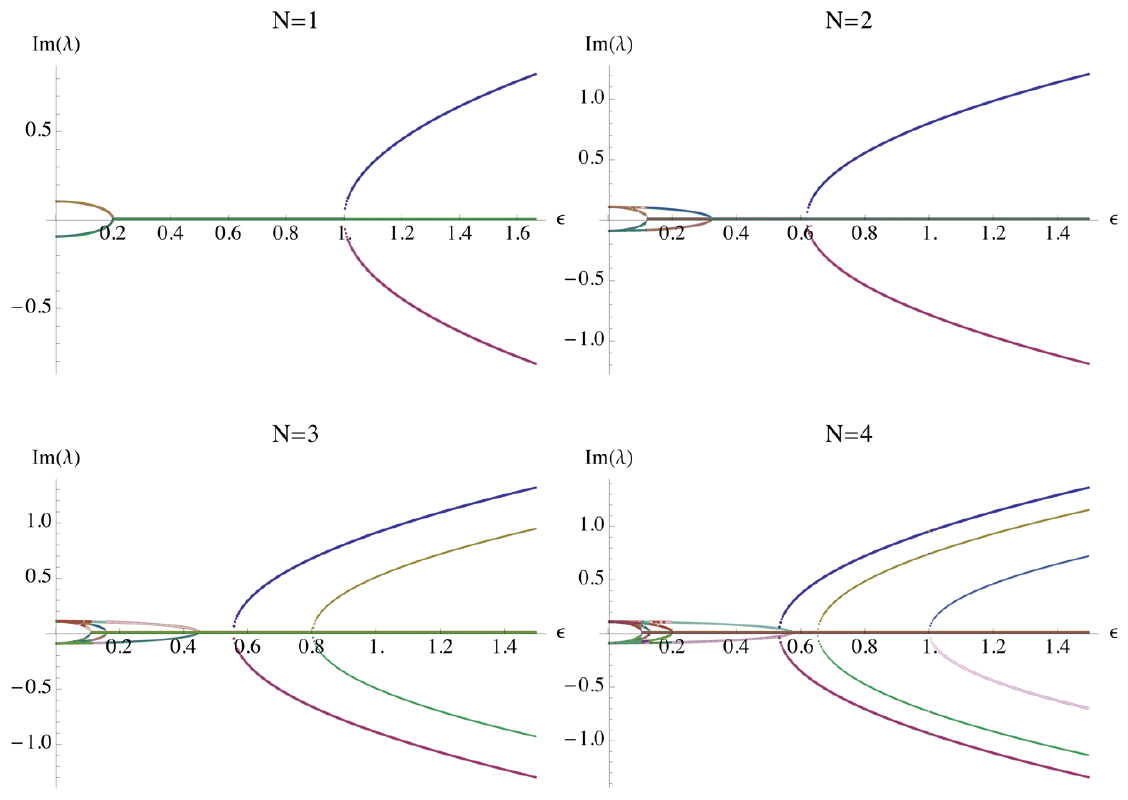}
\end{center}
\caption{Imaginary parts of the frequencies $\lambda$ for $N=1,\,2,\,3,\,$ and
4 as functions of the coupling constant $\epsilon$ for $\omega=1$ and $\gamma=
0.1$. The frequencies are the zeros of the polynomials $P_N$ in (\ref{E32}).
Observe that the extent of the unbroken-$\cPT$-symmetric region (where the
frequencies are all real) decreases as $N$ increases and disappears entirely
when $N=4$.}
\label{F2}
\end{figure}

Given these polynomials, we can calculate the frequencies $\lambda$ to see what
happens to the unbroken-$\cPT$-symmetric region as $N$ increases. In
Fig.~\ref{F2} we plot the imaginary part of $\lambda$ for $N=1,\,2,\,3$, and 4
for fixed $\omega=1$ and $\gamma=0.1$. It is clear that as $N$ increases, the 
size of the unbroken region in the coupling parameter $\epsilon$ shrinks and at
$N=4$ it disappears entirely.

To study analytically the shrinking of the unbroken region with increasing $N$,
we solve the constant-coefficient recursion relation (\ref{E33}). The exact
solution is
\begin{equation}
P_N=\sqrt{\pi}\sum_{k=0}^N(-1)^k\frac{4^{k-N}(2N-k)!}{(N-k)!k!\Gamma(N-k+1/2)}
x^{N-k}\epsilon^{2k}.
\label{E34}
\end{equation}
Substituting $x=-4\epsilon^2 y$ and $\Delta=\sqrt{y(y+1)}$, we express these
polynomials more simply:
\begin{equation}
P_N=\frac{\epsilon^{2N}}{2\Delta}(-1)^N\left[(1+2y-2\Delta)^N(\Delta-y)+(1+2y+2
\Delta)^N(\Delta+y)\right].
\label{E36}
\end{equation}

The zeros of $P_N$ are the roots of the equation $\sqrt{y}+\sqrt{y+1}=(-1)^{1/
(4N+2)}$. Since $y=-\big[\left(\lambda^2-\omega^2\right)^2+4\lambda^2\gamma^2
\big]/\left(4\epsilon^2\right)$ is negative, we substitute $y=-z^2$. The
equation for $z$ then reads $iz+\sqrt{1-z^2}=(-1)^{1/(4N+2)}$, whose solutions
are
\begin{equation}
z=\sin[\pi(2k+1)/(4N+2)]\quad(k=0,1,\ldots,4N+1).
\label{E37}
\end{equation}
Consequently, the equation for $\lambda$ becomes $4\epsilon^2 z^2=(\lambda^2-
\omega^2)^2+4\lambda^2\gamma^2$, whose roots are
\begin{equation}
\lambda_{1,2,3,4}=\pm\sqrt{\omega^2-2\gamma^2\pm2\sqrt{\gamma^2(\gamma^2-
\omega^2)+\epsilon^2z^2}}.
\label{E38}
\end{equation}

We consider two cases. For $N=1$ there are four roots, $z=\sin\theta=\pm1,\,\pm
\half$ with $e^{6i\theta}=-1$. These correspond to the six values $\theta=
\left\{\frac{\pi}{6},\frac{\pi}{2},\frac{5\pi}{6},\frac{7\pi}{6},\frac{3\pi}{2},
\frac{11\pi}{6}\right\}$. The solutions $z=\pm1$ are spurious, and the only
admissable solutions are $z=\pm1/2$. Substituting $z^2=1/4$ into (\ref{E38}), we
obtain the four roots of the polynomial $P_1$ in (\ref{E32}).

For the case $N=2$ there are six roots,
$$z=\sin\theta=\{-1,-(1+\sqrt{5})/4,(1-\sqrt{5})/4,(\sqrt{5}-1)/4,(1+\sqrt{5}
)/4,1\}$$
with $e^{10i\theta}=-1$. These correspond to the ten values
$$\theta=\left\{\frac{1}{10}\pi,\frac{3}{10}\pi,\frac{5}{10}\pi,\frac{7}{10}\pi,
\frac{9}{10}\pi,\frac{11}{10}\pi,\frac{13}{10}\pi,\frac{3}{2}\pi,\frac{17}{10}
\pi,\frac{19}{10}\pi\right\}.$$
The solutions $z=\pm1$ are spurious and there are only four genuine roots $z=\pm
(1\pm\sqrt{5})/4$ and two values $z^2=(1\pm\sqrt{5})^2/16$ to substitute into
(\ref{E38}) for getting the eight roots of the polynomial $P_2$ in (\ref{E32}).

In general, in the region of unbroken $\cPT$ symmetry the roots $\lambda$ in
(\ref{E38}) are all real. Thus,
\begin{equation}
0<\gamma<\sqrt{\omega^2/2-\sqrt{\omega^4/4-\epsilon^2z_{\rm min}^2}},\quad\gamma
\sqrt{\omega^2-\gamma^2}/\sqrt{z_{\rm min}}<\epsilon< \omega^2/(2z_{\rm max}),
\label{E39}
\end{equation}
where $z_{\rm min}=\sin[\pi/(4N+1)]$ and $z_{\rm max}=\sin[\pi(2N-1)/(4N+1)]$.
Condition (\ref{E39}) identifies the region in the parameter space $(\gamma,
\epsilon)$ where the $\cPT$ symmetry is unbroken. Note that as $N\to\infty$,
$z_{\rm min}\to0$ and $z_{\rm max}\to1$. Thus, as $N\to\infty$ the only allowed
$\gamma$ is 0 (so that there is no loss and gain), and the range of $\epsilon$
shrinks to $0\leq\epsilon<\omega^2/2$.

\begin{figure}[h!]
\begin{center}
\includegraphics[scale=1.50]{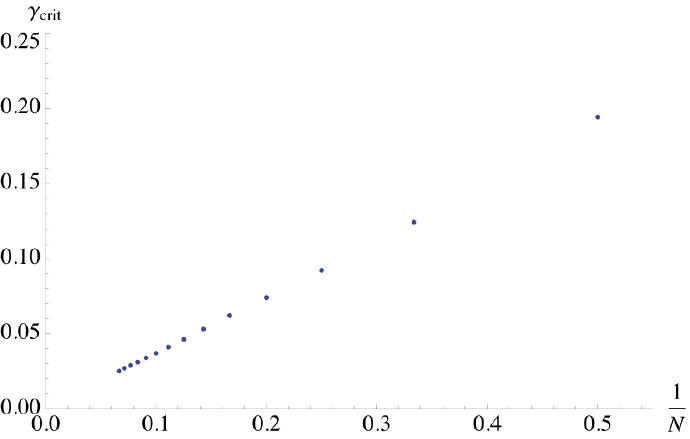}
\end{center}
\caption{Plot of $\gamma_{\rm crit}$ as a function of $1/N$. This sequence
evidently converges to 0 with increasing $N$. Thus, a system of coupled
oscillators with a uniform loss-gain parameter $\gamma>0$ has no unbroken
$\cPT$-symmetric region if $N$ is sufficiently large.}
\label{F3}
\end{figure}

Let us examine further how the allowed $\gamma$ decreases as a function of
increasing $N$. We can see from Fig.~\ref{F2} that at the lower end of the
unbroken region the curves open to the left and at the upper end of this region
the curves open to the right. For fixed $N$ and fixed $\epsilon=\omega^2/(2z_{
\rm max})$, if we increase $\gamma$, the left opening curves will eventually
touch the right opening curves and the unbroken region in $\epsilon$ will
disappear. We designate as $\gamma_{\rm crit}$ the critical value of $\gamma$ at
which the unbroken region in $\epsilon$ disappears. If we compute $\gamma_{\rm
crit}$ as a function of $N$ and plot in Fig.~\ref{F3} these values of $\gamma_{
\rm crit}$ versus $1/N$, we see clearly that the critical value of $\gamma$
decreases to 0. Thus, if there are too many oscillators, there cannot be a
region of unbroken $\cPT$ symmetry in a system with uniform nonzero loss and
gain.

\begin{figure}[h!]
\begin{center}
\includegraphics[scale=0.75]{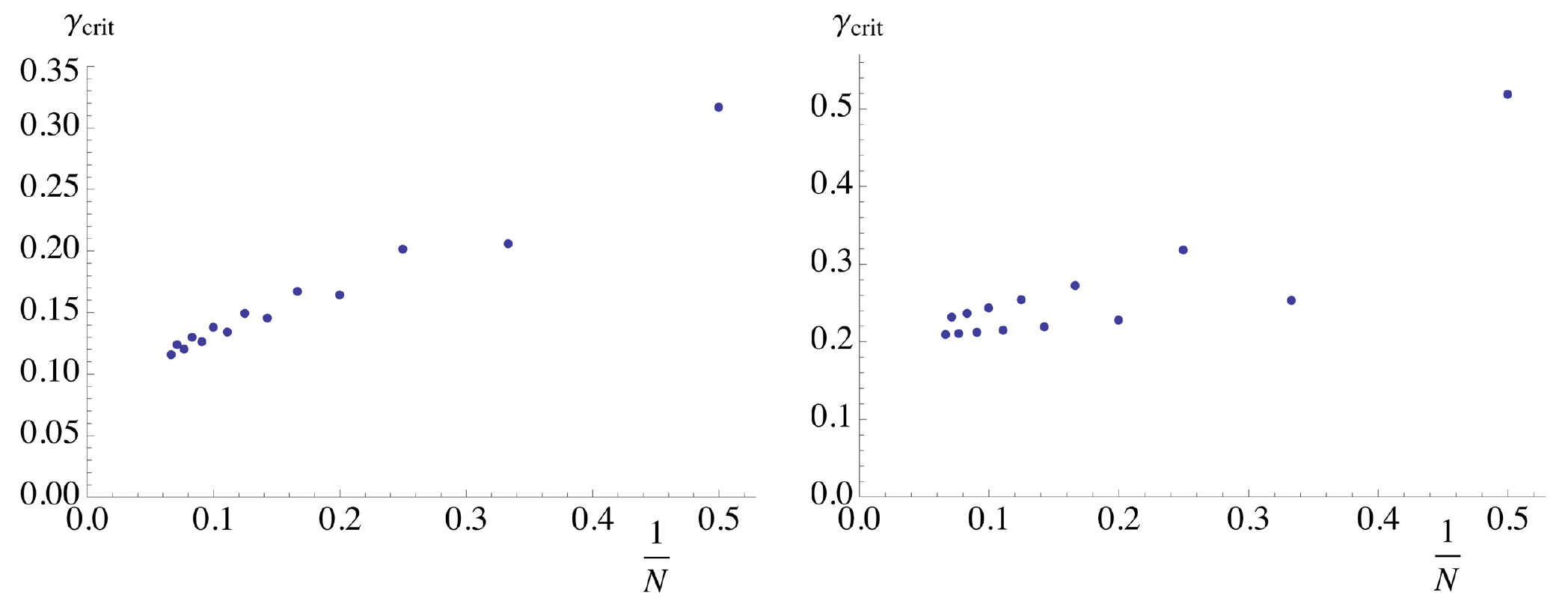}
\end{center}
\caption{Analog of Fig.~\ref{F3}: Oscillatory convergence of $\gamma_{\rm crit}$
when the
loss-gain parameter $\gamma_n$ decreases like $\gamma/n$ (left panel) and like
$\gamma/n^2$ (right panel). Evidently, if the loss-gain parameter decays to zero
for more distant oscillators, a region of unbroken $\cPT$ symmetry can persist
as $N\to\infty$.}
\label{F4}
\end{figure}

The only way for an unbroken region of $\cPT$ symmetry to survive as $N\to
\infty$ is for the loss-gain parameter to decrease with increasingly distant
oscillators. Our numerical calculations show that if the loss-gain parameter is
$\gamma/(N-n+1)$ (where $n$ ranges from 1 to $N$), there will be an unbroken
region if $\gamma$ is less than about $0.1$ (Fig.~\ref{F4}, left panel), and
if the loss-gain parameter is $\gamma/(N-n+1)^2$ (where $n$ ranges from 1 to
$N$), there will be an unbroken region if $\gamma$ is less than about $0.2$
(Fig.~\ref{F4}, right panel).

\section{Localized impurity in the continuum model}
\label{s5}
In Sec.~\ref{s4} we showed that if the effect of loss and gain is localized
about the central oscillators and decays for more distant oscillators, then the
unbroken-$\cPT$-symmetric region can survive as $N\to\infty$. This suggests that
for the continuum model developed in subsection \ref{ss3A} it would be
interesting to examine what happens when $\gamma(x)$ decreases with increasing
$|x|$. The simplest case to study is that for which $\gamma(x)=\gamma\delta(x)$;
that is, the case of a localized point-like $\cPT$-symmetric loss-gain impurity
at the origin. Studies of this type have been performed for tight-binding models
by Joglekar {\it et al} \cite{R17,R18} and Longhi \cite{R19}.

Let us assume that the loss-gain parameter is a localized function of $x$ at the
origin, $\gamma(x)=\gamma\delta(x)$, and seek a solution to (\ref{E30}) with
frequency $\Omega$:
\begin{equation}
S(x,t)=e^{i\Omega t}s(x),\quad D(x,t)=e^{i\Omega t}d(x).
\label{E40}
\end{equation}
If we assume that $a^2=\omega^2-\Omega^2+\epsilon>0$ and that $\quad-b^2=
\omega^2-\Omega^2-\epsilon<0$, where $a$ and $b$ are positive, the coupled wave
equations become coupled ordinary differential equations:
\begin{equation}
c^2s''(x)-a^2 s(x)=2i\Omega\gamma\delta(x)d(x)\quad{\rm and}\quad
c^2d''(x)+b^2 d(x)=2i\Omega\gamma\delta(x)s(x).
\label{E41}
\end{equation}
The functions $s(x)$ and $d(x)$ are continuous at $x=0$ and the delta function
gives rise to a discontinuity in the derivatives of $s$ and $d$ at $x=0$:
\begin{eqnarray}
2i\gamma\Omega d(0)=c^2\big[s'(0^+)-s'(0^-)\big]\quad{\rm and}\quad
2i\gamma\Omega s(0)=c^2\big[d'(0^+)-d'(0^-)\big].
\label{E42}
\end{eqnarray}

A simple solution to (\ref{E41}) has the form
\begin{eqnarray}
s(x)=e^{-a|x|/c}\quad{\rm and}\quad d(x)&=&i\frac{ac}{\gamma\Omega}\cos\frac{bx}
{c}+i\frac{\gamma\Omega}{bc}\sin\frac{b|x|}{c}.
\label{E43}
\end{eqnarray}
This solution is $\cPT$ symmetric, where $\cP$ changes the sign of $x$ and
interchanges $u$ and $v$, which in turn changes the sign of $d$ while leaving
the sign of $s$ unchanged, and $\cT$ performs complex conjugation.

This solution can be viewed as a pseudo-bound-state solution in the sense that
$s(x)$ decays exponentially as $|x|\to\infty$. However, while $d(x)$ also has a
cusp at $x=0$, it is not localized and oscillates as $|x|\to\infty$. This
solution resembles that found by Hatano {\it et al} \cite{R20,R21} and Longhi
\cite{R19}. It is interesting that no localized bound-state solution exists if 
$a^2=\omega^2-\Omega^2+\epsilon>0$ and $b^2=\omega^2-\Omega^2-\epsilon>0$, where
$a$ and $b$ are positive.

\section{Three planar oscillators}
\label{s6}
It appears that for all one-dimensional chains of oscillators there is just one
region of unbroken $\cPT$ symmetry. However, it is possible to have more than
one region of unbroken $\cPT$ symmetry if the oscillators are coupled in a
planar array. For example, let us consider three oscillators in a plane, where
the first (the $x$ oscillator) has loss, the second (the $y$ oscillator) has
gain, and the third (the $z$ oscillator) has neither loss nor gain. The $x$ and
$y$ oscillators are coupled directly and are also coupled indirectly through the
$z$ oscillator. The Hamiltonian for this system is
\begin{eqnarray}
H=\frac{\omega_2^2}{4}q^2+\frac{\omega_2^2}{2}pr+y^2+2\frac{\omega_1^2-\gamma^2}
{\omega_2^2}xz-2\frac{\epsilon_1}{\omega_2^2}(xy+yz)-\frac{\epsilon_2}{
\omega_2^2}(x^2+z^2)+\gamma(zr-xp).
\label{E44}
\end{eqnarray}
This Hamiltonian gives the equations of motion
\begin{equation}
\ddot{x}+\omega_1^2 x+2\gamma\dot{x}=\epsilon_1 y+\epsilon_2 z,\quad
\ddot{y}+\omega_2^2 y=\epsilon_1(x+z),\quad
\ddot{z}+\omega_1^2 z-2\gamma \dot{z}=\epsilon_1 y+\epsilon_2 x.
\label{E45}
\end{equation}

This oscillator system can have {\it two} regions of unbroken $\cPT$ symmetry.
Without loss of generality, we choose $\omega_2=1$ and $\omega_1=\omega$ so that
$H$ in (\ref{E44}) becomes
\begin{equation}
H=\frac{1}{4}q^2+\frac{1}{2}pr+y^2+2(\omega^2-\gamma^2)xz-2\epsilon_1(xy+yz)-
\epsilon_2(x^2+z^2)+\gamma(zr-xp)
\label{E46}
\end{equation}
and the system of equations (\ref{E45}) becomes
\begin{equation}
\ddot{x}+\omega^2 x+2\gamma \dot{x}=\epsilon_1y+\epsilon_2z,\quad\ddot{y}+y=
\epsilon_1(x+z),\quad\ddot{z}+\omega^2 z-2\gamma\dot{z}=\epsilon_1y+\epsilon_2x.
\label{E47}
\end{equation}
To find the frequencies of this classical system, we seek solutions to
(\ref{E47}) of the form $x(t)=Ae^{i\lambda t},\,y(t)=Be^{i\lambda t},\,z(t)=C
e^{i\lambda t}$. We use Cramer's rule to eliminate the coefficients $A$, $B$,
and $C$, and find that the resulting equation for the frequency $\lambda$ is
$$P(\lambda)=\lambda^6+\lambda^4(4\gamma^2-2\omega^2-1)+\lambda^2(\omega^4+2
\omega^2-2\epsilon_1^2-\epsilon_2^2-4\gamma^2)+2\epsilon_1^2(\epsilon_2+\omega^2
)+\epsilon_2^2-\omega^4.$$

\begin{figure}[h!]
\begin{center}
\includegraphics[scale=0.36]{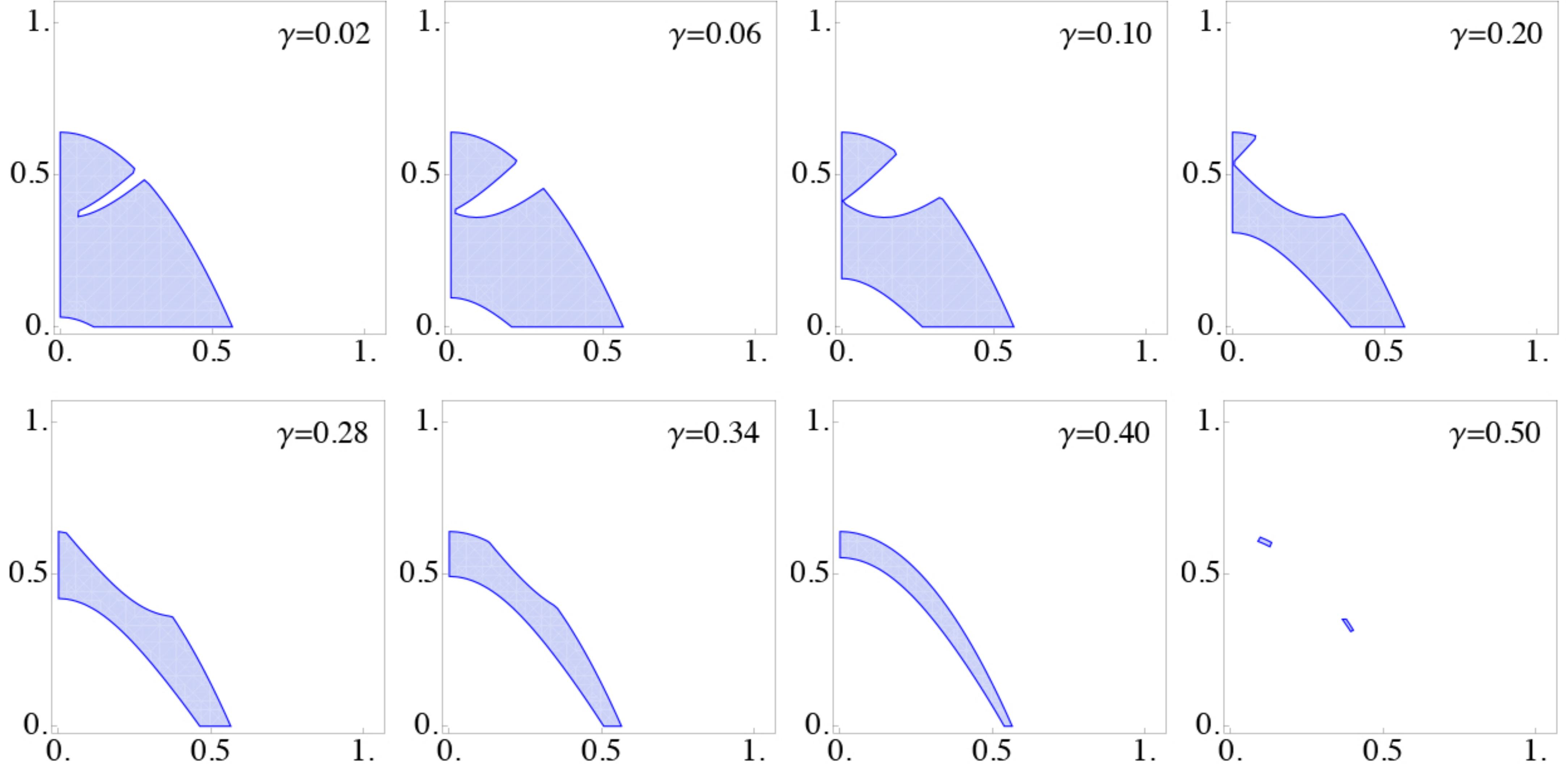}
\end{center}
\caption{Regions in the space of parameters $(\epsilon_1~[{\rm
horizontal~axis}],\epsilon_2~[{\rm vertical~axis}])$ for which the $\cPT$
symmetry is unbroken; that is, the roots of $P(\lambda)$ in (\ref{E48}) are all
real and positive. For this figure the frequency $\omega=0.8$ and the damping
parameter has the values $\gamma=0.02,\,0.06,\,0.10,\,0.20,\,0.28,\,0.34,\,0.40,
\,0.50$. As $\gamma$ increases, the unbroken-$\cPT$-symmetric regions in
$(\epsilon_1,\epsilon_2)$ space decrease in size and eventually disappear.
Unlike the case of linear chains of $\cPT$-symmetric coupled oscillators,
as $\epsilon_2$ increases from $0$ for fixed $\epsilon_1$, there is a range of
$\gamma$ and $\omega$ such that one can observe {\it five} regions of broken,
unbroken, broken, unbroken, and broken $\cPT$ symmetry. For example, there are
five regions when $\gamma=0.10$, $\epsilon_1=0.10$, and $0\leq\epsilon_2\leq
0.70$.}
\label{F5}
\end{figure}

\begin{figure}[h!]
\begin{center}
\includegraphics[scale=0.36]{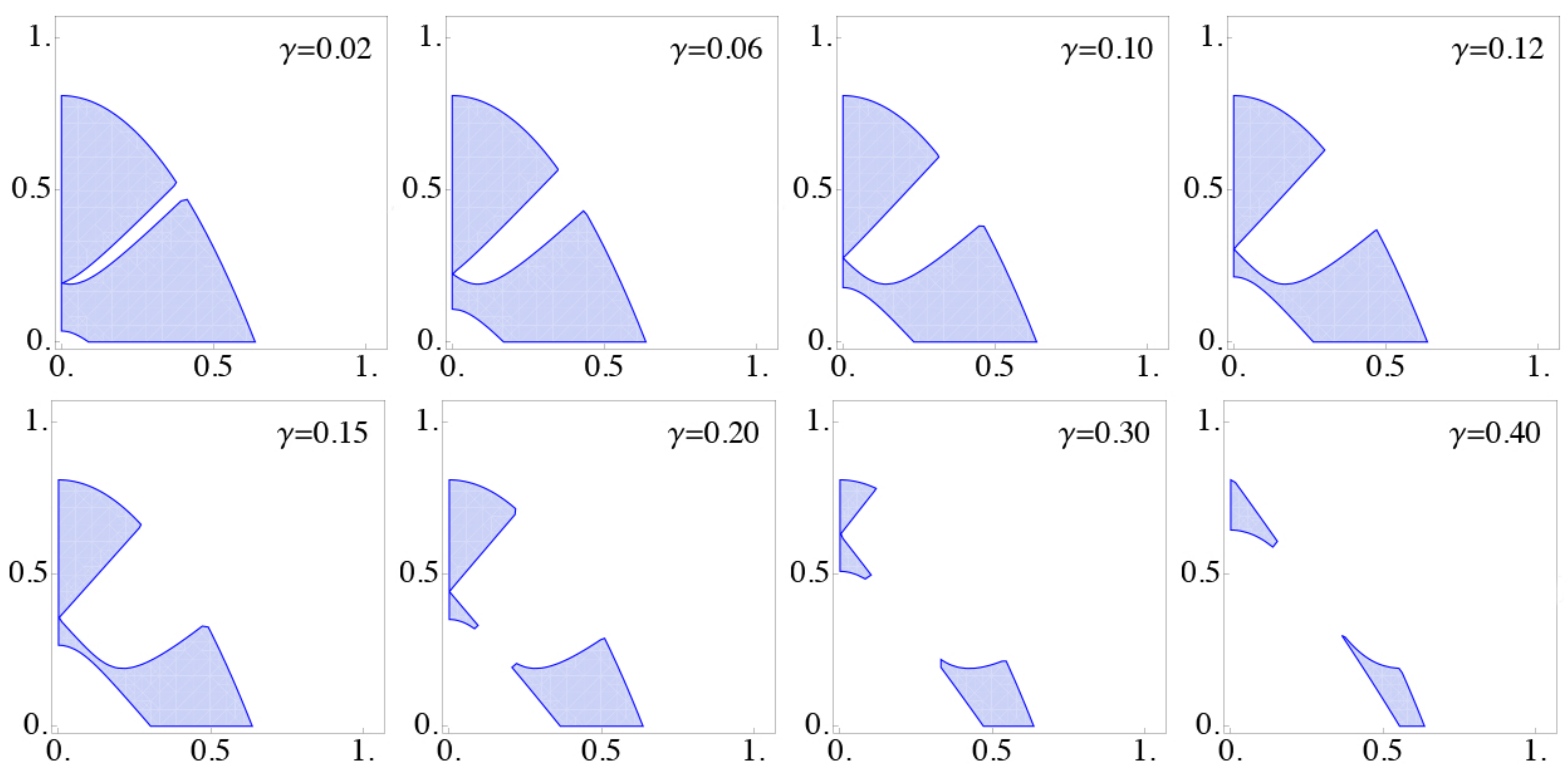}
\end{center}
\caption{Same as in Fig.~\ref{F5} but with $\omega=0.9$.}
\label{F6}
\end{figure}

\begin{figure}[h!]
\begin{center}
\includegraphics[scale=0.36]{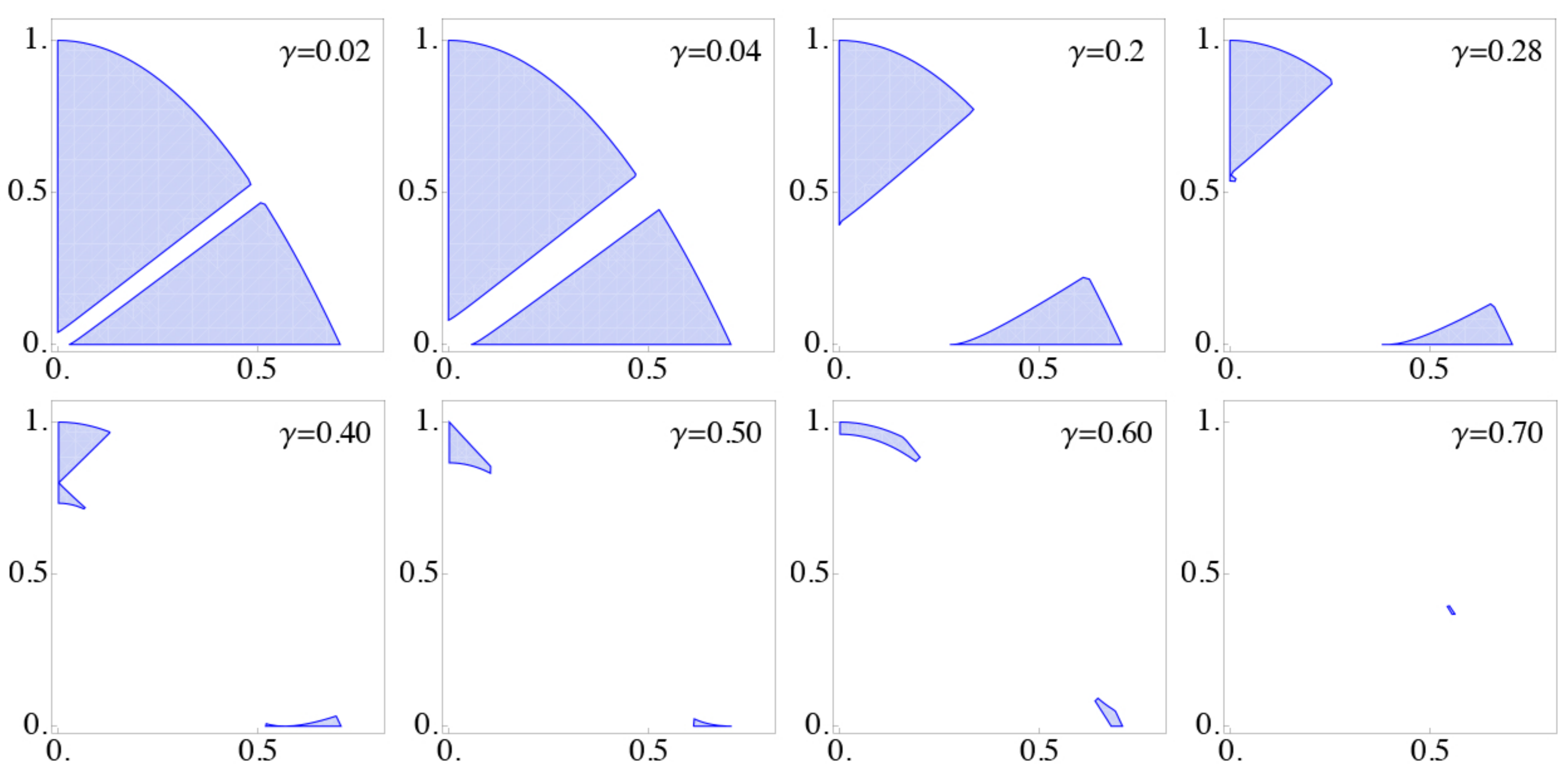}
\end{center}
\caption{Same as in Fig.~\ref{F5} but with $\omega=1.0$.}
\label{F7}
\end{figure}

\begin{figure}[h!]
\begin{center}
\includegraphics[scale=0.36]{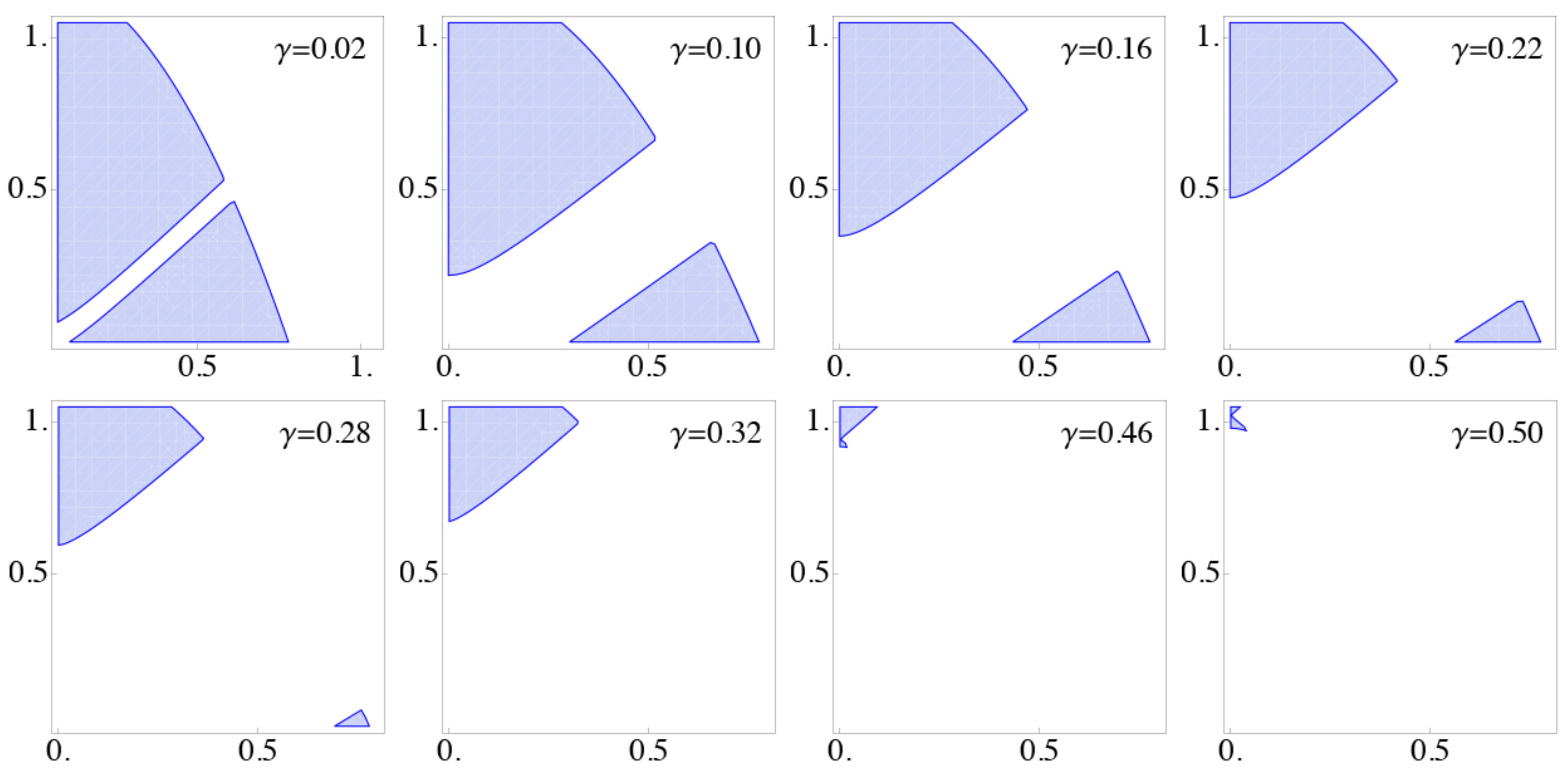}
\end{center}
\caption{Same as in Fig.~\ref{F5} but with $\omega=1.1$.}
\label{F8}
\end{figure}

\begin{figure}[h!]
\begin{center}
\includegraphics[scale=0.52]{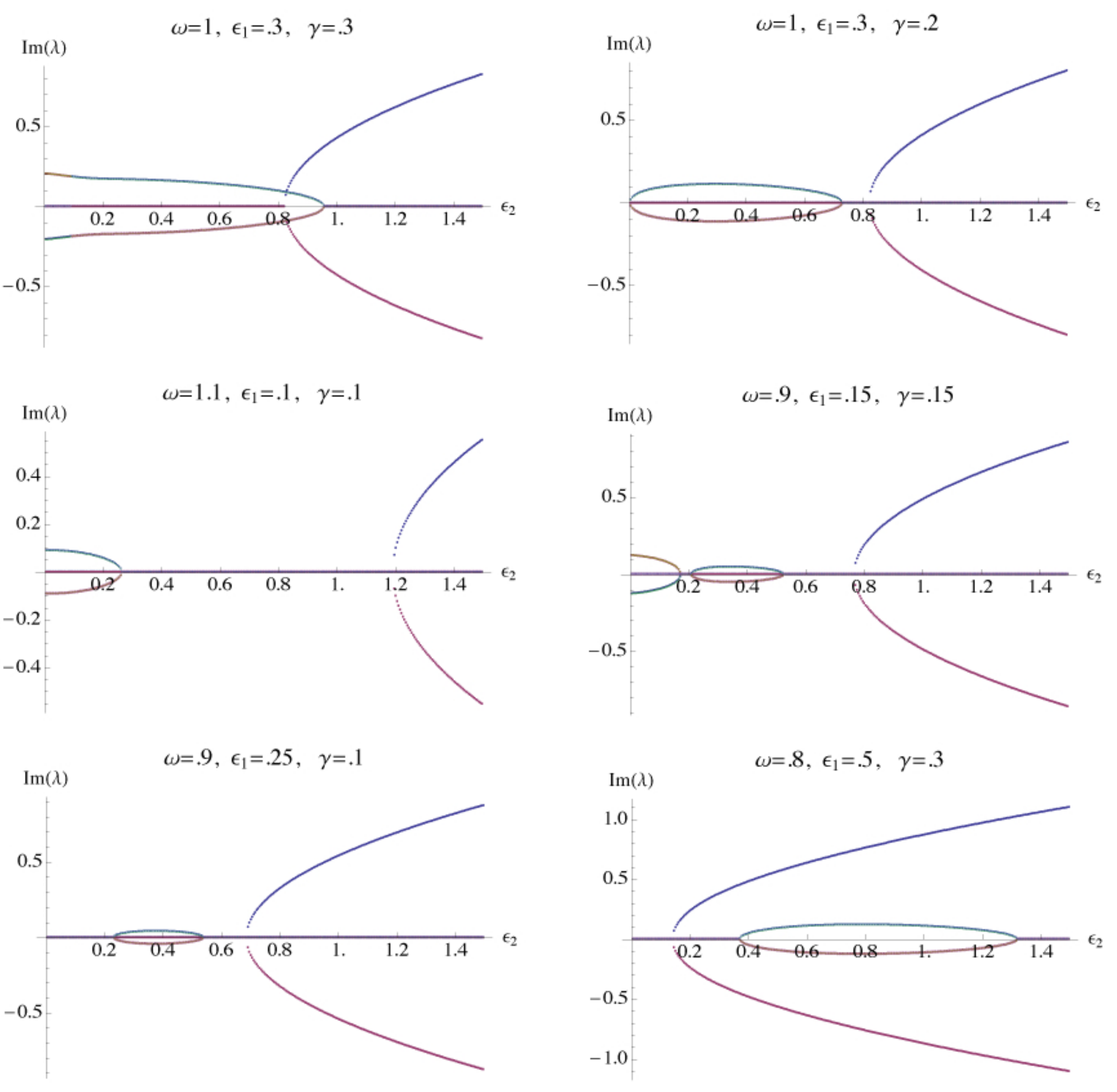}
\end{center}
\caption{Imaginary parts of the frequencies $\lambda$ plotted as a function of
$\epsilon_2$ for various values of the parameters $\omega$, $\gamma$, and
$\epsilon_1$. The regions of unbroken $\cPT$ symmetry occur when the imaginary
parts vanish and all frequencies are real.}
\label{F9}
\end{figure}

With the substitution $\mu=\lambda^2$, this polynomial becomes
\begin{equation}
p(\mu)=\mu^3-\alpha\mu^2+\beta\mu-\sigma
\label{E48}
\end{equation}
with coefficients $\alpha=1+2\omega^2-4\gamma^2$, $\beta=\omega^4+2\omega^2-2
\epsilon_1^2-\epsilon_2^2-4\gamma^2$, $\sigma=\omega^4-2\epsilon_1^2(\epsilon_2+
\omega^2)-\epsilon_2^2$. Positive real roots of (\ref{E48}) are obtained by
searching for the regions in the parameter space where the minimum $\mu_m=
(\alpha-\sqrt{\alpha^2-3\beta})/3$ and maximum $\mu_p=(\alpha+\sqrt{\alpha^2-3
\beta})/3$ are real and positive, and $p(\mu_m)>0$ and $p(\mu_p)<0$.
Figures~\ref{F5}, \ref{F6}, \ref{F7}, and \ref{F8} display the regions of
unbroken $\cPT$ symmetry [where the roots of $P(\lambda)$ in (\ref{E48}) are all
real] for various values of $\omega$, $\gamma$, $\epsilon_1$, and $\epsilon_2$.
For special ranges of the parameters $\omega$, $\gamma$, and $\epsilon_1$ one
can get five distinct regions of broken and unbroken $\cPT$ symmetry as
$\epsilon_2$ increases continuously from $0$. (For the case of linear chains of
$\cPT$-symmetric coupled oscillators one can have at most three regions.) The
imaginary parts of the frequencies $\lambda$ as functions of $\epsilon_2$ are
plotted in Fig.~\ref{F9}. The unbroken-$\cPT$-symmetric regions are
characterized by the vanishing of ${\rm Im}\,\lambda$.

\section{Brief concluding remarks}
\label{s7}
The purpose of this paper has been to examine physically constructable
$\cPT$-symmetric systems consisting of many coupled oscillators. (Similar
studies have been done for $\cPT$-symmetric arrays of optical waveguides with
loss and gain \cite{R22,R23}.) We have implemented $\cPT$-symmetry by arranging
the oscillators so that loss and gain are balanced pairwise. We have examined
one-dimensional systems consisting of both even and odd numbers of oscillators,
and have also studied the limiting behavior as the number of oscillators
approaches infinity. We have shown that the Hamiltonians associated with these
systems can be formulated in two different ways, first as a sum representation
and second as a product representation. The latter representation has a 
gauge-like coupling structure that can be used to demonstrate that the
Hamiltonian is not unique.

We have shown that when the oscillators are arranged in a one-dimensional chain,
for sufficiently many oscillators there cannot be a region of unbroken $\cPT$
symmetry (where the frequencies are all real) unless the loss-gain parameter
$\gamma$ decays with the distance from the center of the chain. Our numerical
calculations show that if $\gamma$ decays fast enough, then a region of unbroken
$\cPT$ symmetry will always exist, even if the number of oscillators is
infinite. We have also shown that in the continuum limit, a localized gain-loss
impurity can give rise to a pseudo-bound state.

Our analysis shows that for a one-dimensional chain of oscillators, as the
coupling constant $\epsilon$ increases from $0$, one can find at most only three
regions, two regions of broken $\cPT$ symmetry surrounding a region of unbroken
$\cPT$ symmetry. However, a two-dimensional array of oscillators can exhibit
more than three regions. For example, a triangle of coupled oscillators can
exhibit five regions. Optics experiments are currently underway to study such a
system \cite{R24}.

\acknowledgments
CMB is grateful for the hospitality of the Heidelberg Graduate School of
Fundamental Physics. CMB thanks the U.S.~Department of Energy and MG thanks the
Fondazione Angelo Della Riccia for financial support.


\begin{thebibliography}{99}
\bibitem{R2} The minus signs in (\ref{E6}) are optional; they are included so
that this definition of parity reduces to the conventional one for the case of
one oscillator.
\bibitem{R16} There smay be an interesting connection with the Calogero model,
which is also a gauge theory with $N$ degrees of freedom. See, for example,
A.~P.~Polychronakos, J.~Phys.~A: Math.~Gen.~{\bf 39}, 12793 (2006).
\end{thebibliography}
\end{document}